\documentclass[reprint, twocolumn, superscriptaddress,prb]{revtex4-2}
\usepackage{bm, amsmath, amsfonts, amssymb, braket}
\usepackage{subfigure}
\usepackage{color}
\usepackage{graphicx}
\usepackage{slashed}

\usepackage{dcolumn} 

\usepackage{rotating} 

\usepackage{ulem}
\usepackage{xcolor}
\DeclareRobustCommand{\erase}{\bgroup\markoverwith{\textcolor{red}{\rule[.5ex]{2pt}{0.4pt}}}\ULon}

\begin{document}
\title{Predicting quantum ground-state energy by data-driven Koopman analysis of variational parameter nonlinear dynamics}
\author{Nobuyuki Okuma}
\email{okuma@hosi.phys.s.u-tokyo.ac.jp}

\affiliation{%
  Graduate School of Engineering, Kyushu Institute of Technology, Kitakyushu 804-8550, Japan
}%

\date{\today}
\begin{abstract}
In recent years, the application of machine learning to physics has been actively explored. In this paper, we study a method for estimating the ground-state energy of quantum Hamiltonians by applying data-driven Koopman analysis within the framework of variational wave functions.
Koopman theory is a framework for analyzing the nonlinear dynamics of vectors, in which the dynamics are linearized by lifting the vectors to functions defined over the original vector space. We focus on the fact that the imaginary-time Schrödinger equation, when restricted to a variational wave function, is described by a nonlinear time evolution of the variational parameter vector.
We collect sample points of this nonlinear dynamics at parameter configurations where the discrepancy between the true imaginary-time dynamics and the dynamics on the variational manifold is small, and perform data-driven continuous Koopman analysis. Within our formulation, the ground-state energy is reduced to the leading eigenvalue of a differential operator known as the Koopman generator.
As a concrete example, we generate samples for the four-site transverse-field Ising model and estimate the ground-state energy using extended dynamic mode decomposition (EDMD). Furthermore, as an extension of this framework, we formulate the method for the case where the variational wave function is given by a uniform matrix product state on an infinite chain.
By employing computational techniques developed within the framework of the time-dependent variational principle, all the quantities required for our analysis—including error estimation—can be computed efficiently in such systems. Since our approach provides predictions for the ground-state energy even when the true ground state lies outside the variational manifold, it is expected to complement conventional variational methods.

\end{abstract}
\maketitle
\section{Introduction}
Machine learning (ML) has been applied to a wide range of problems in physics in recent years. One direction is physics-informed learning \cite{raissi2019physics}, where known physical structures such as differential equations, conservation laws, or symmetries are incorporated into the model to guide training. Another line of work uses ML for the identification of phases of matter \cite{carrasquilla2017machine,wang2016discovering,van2017learning}, where supervised or unsupervised methods are employed to classify different phases and detect phase transitions from data. In addition, ML has been used to represent quantum many-body wave functions \cite{carleo2017solving}, for example, by constructing variational ansätze based on neural networks and optimizing them with respect to physical observables. While these approaches differ in their objectives and methodologies, they illustrate the growing role of ML as a flexible tool for tackling diverse problems in physics \cite{carleo2019machine,mehta2019high}.

In this paper, we apply data-driven Koopman analysis \cite{koopman1931hamiltonian,koopman1932dynamical,schmid2010dynamic, williams2015data,klus2020data,brunton2021modern} to estimate the ground-state energy of quantum many-body systems. Koopman theory is one of the methods for analyzing the nonlinear time evolution of vectors \cite{koopman1931hamiltonian,koopman1932dynamical}. Within this framework, vectors are lifted to functions—i.e., to infinite-dimensional vectors—and the given nonlinear dynamics are represented as a linear time evolution defined by an operator called the Koopman operator (or its generator). In other words, by paying the cost of moving to an infinite-dimensional space, one can linearize a nonlinear problem. This framework is not only of theoretical interest but has also attracted significant attention in recent years as a data-driven approach \cite{schmid2010dynamic, williams2015data,klus2020data,brunton2021modern}.
In this study, we consider the imaginary-time Schrödinger equation, which describes a linear time evolution governed by a large but finite-dimensional Hamiltonian. As a first step, we reduce this linear time evolution to a low-dimensional nonlinear dynamical system. Specifically, we exploit the fact that the imaginary-time evolution within the manifold spanned by a variational wave function can be reduced to a nonlinear time evolution of the variational parameter vector. 
Focusing on this point, we analyze the nonlinear dynamical data of variational parameters using extended dynamic mode decomposition (EDMD) \cite{williams2015data,klus2020data}.
Within the formulation of this paper, the ground-state energy of the original quantum Hamiltonian is given by the leading eigenvalue of the Koopman generator.
By this approach, a linear equation for a complicated many-body system, which is difficult to handle directly, is mapped—via a nonlinear system—into a linear equation that is more amenable to ML.
As an application, we also formulate the method for the case where the variational wave function is given by a uniform matrix product state on an infinite chain.

This paper is organized as follows. In Sec. \ref{sec2}, we establish a connection between the imaginary-time Schrödinger equation restricted to the variational space, the resulting nonlinear time evolution, and Koopman analysis. In this context, the ground-state energy of the quantum Hamiltonian corresponds to the leading eigenvalue of the Koopman generator.
In Sec. \ref{sec3}, we investigate the properties of Koopman eigenvalues and eigenfunctions using an analytically tractable two-level system as an example.
In Sec. \ref{sec4}, we numerically study the four-site transverse-field Ising model as a minimal example in which the imaginary-time dynamics does not close within the variational space. Using samples drawn from the variational space, we perform a preliminary analysis based on EDMD \cite{williams2015data, klus2020data} and estimate the ground-state energy.
In Sec. \ref{section:mps}, as an application of the present method, we formulate the approach for the case of matrix product states on an infinite chain.
In such systems, all the quantities required for our Koopman analysis are efficiently computed within the framework of the time-dependent variational principle for matrix product states \cite{haegeman2011time,haegeman2013post,haegeman2016unifying,vanderstraeten2019tangent}.
In Sec. \ref{discussion}, we discuss the remaining issues and outline directions for future work.

\section{Formalism\label{sec2}}
In this section, we establish a connection between the imaginary-time Schrödinger equation under restriction to the variational space, the resulting nonlinear time evolution, and Koopman analysis.

\subsection{Koopman theory for variational-parameter nonlinear dynamics}
The imaginary-time Schr\"{o}dinger equation without imposing norm conservation is given by
\begin{align}
    \frac{d}{d\tau}\ket{\psi}=-H\ket{\psi},\label{imgeqn}
\end{align}
where $H$ is a $D\times D$ Hermitian matrix representation of the quantum Hamiltonian, and $\ket{\psi}\in\mathbb{C}^D$ is a non-normalized state vector. 
In this paper, we assume that the zero of energy is chosen such that the ground-state energy is positive.
Thus, as $\tau\rightarrow\infty$, $\ket{\psi}$ becomes the zero vector.
If the state vector has a nonzero overlap with the ground states, then after a sufficiently long time,
$\ket{\psi}$ becomes proportional to one of the ground states.
In the latter part of the paper, we also consider the imaginary-time Schr\"{o}dinger equation including a norm-conservation term, aiming at applications to various many-body methods:
\begin{align}
    \frac{d}{d\tau}\ket{\phi}=-\left(H-\langle H\rangle_{\phi}\right)\ket{\phi},\label{app2}
\end{align}
where $\ket{\phi}$ is a normalized state, and $\langle H\rangle_{\phi}:=\bra{\phi}H\ket{\phi}$.

In the following, we study the above dynamics within the space of variational wave functions $\ket{\psi_{\bm{\theta}}}$, where $\bm{\theta}$ is the parameter vector.
First, we assume that the variational space is sufficiently large, and the imaginary-time dynamics is always closed within the variational space:
\begin{align}
    &\frac{d}{d\tau}\ket{\psi_{\bm{\theta}}}=-H\ket{\psi_{\bm{\theta}}}\notag\\
    \Leftrightarrow&\dot{\bm{\theta}}\cdot\nabla_{\bm{\theta}}\ket{\psi_{\bm{\theta}}}=-H\ket{\psi_{\bm{\theta}}},\label{variationalsch}
\end{align}
where $\dot{\bm{\theta}}:=d\bm{\theta}/d\tau$.
In practice, this assumption is too restrictive, and we will relax it later.
Since the left-hand side depends only on $\bm{\theta}$, $\dot{\bm{\theta}}$ depends only on $\bm{\theta}$.
Thus, we can define the following nonlinear dynamics:
\begin{align}
    \dot{\bm{\theta}}=\bm{f}(\bm{\theta}).\label{nonlinear}
\end{align}
In this nonlinear dynamics, the physics of the original system defined by the variational space—such as information about low-energy states—is extracted.

From a different perspective, by lifting through the map $\bm{\theta}\rightarrow \ket{\psi_{\bm{\theta}}}$,
the nonlinear dynamics of $\bm{\theta}$ in Eq. (\ref{nonlinear}) is transformed into a linear differential equation:
\begin{align}
   \bm{f}(\bm{\theta})\cdot\nabla_{\bm{\theta}}\ket{\psi_{\bm{\theta}}}=-H\ket{\psi_{\bm{\theta}}}.\label{finite-rep}
\end{align}
Expanding in the eigenstates $\{\ket{i}\}$ of $H$, we obtain
\begin{align}
    \bm{f}(\bm{\theta})\cdot\nabla_{\bm{\theta}}~z_i(\bm{\theta})=-E_i~z_i(\bm{\theta}),\label{koopmaneigen}
\end{align}
where $E_i$ is the $i$-th eigenvalue, and $z_i(\bm{\theta}):=\bra{i} \psi_{\bm{\theta}}\rangle$.

\begin{figure}[]
\begin{center}
 \includegraphics[width=8cm,angle=0,clip]{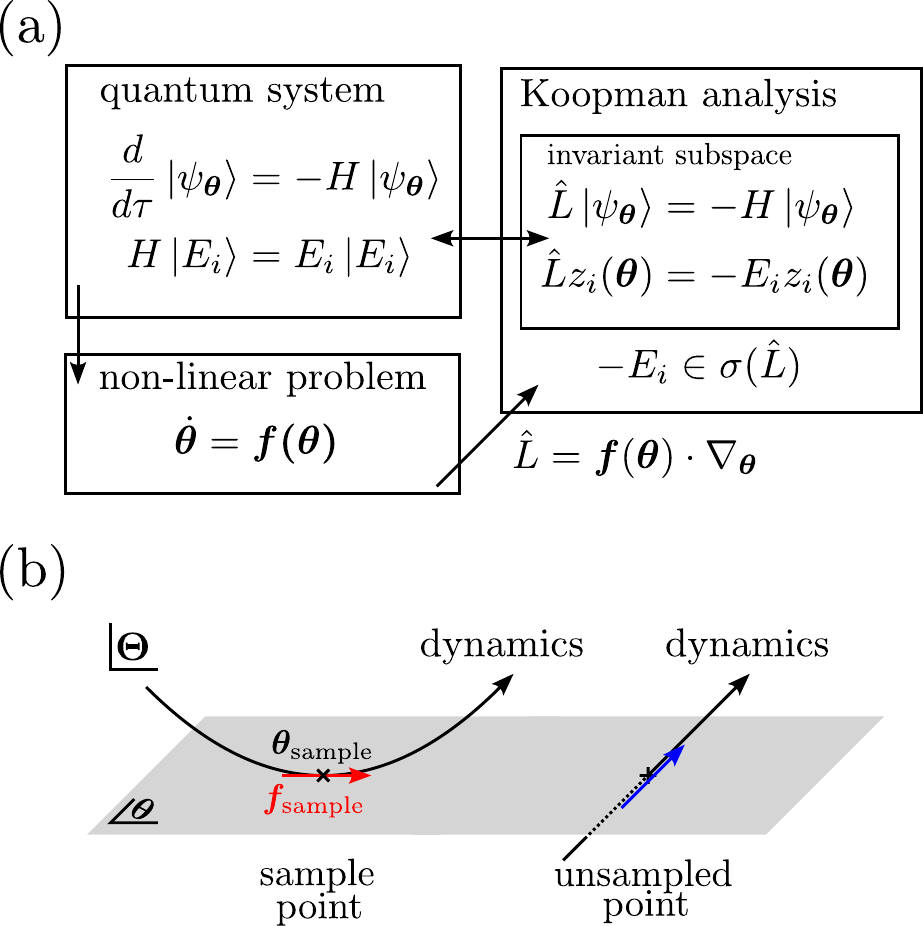}
 \caption{(a) Relationships among quantum system, nonlinear problem, and Koopman analysis. (b) Schematic picture of a sample point.}
 \label{fig1}
\end{center}
\end{figure}

The above discussion can be reinterpreted in terms of Koopman theory \cite{koopman1931hamiltonian,koopman1932dynamical}.
In general, analyzing the nonlinear time evolution of a vector is not straightforward. Koopman theory is a framework that lifts the equations of a finite-dimensional vector to an infinite-dimensional vector—that is, a function—and thereby represents them as a linear equation. In other words, it linearizes the nonlinear problem at the cost of moving to an infinite-dimensional space.
The linear operator $\hat{L}=\bm{f}(\bm{\theta})\cdot\nabla_{\bm{\theta}}$ is called the Koopman generator of the nonlinear dynamics (\ref{nonlinear}).
In Koopman theory, the dynamics of the original nonlinear system are represented through a Koopman mode decomposition based on the eigenvalues and eigenfunctions of the Koopman generator \cite{mezic2005spectral,rowley2009spectral}; therefore, spectral analysis of the generator becomes a central task.
In our problem, since $-E_i$ and $z_i(\bm{\theta})$ correspond to the eigenvalues and eigenfunctions of the Koopman generator, respectively, methods developed in Koopman theory can be used to compute the eigenenergies of the original quantum system.
Note that, in general, $\{E_i\}$ forms only a subset of the spectrum of $\hat{L}$.
In our construction, the Koopman generator admits a nontrivial finite-dimensional invariant subspace, where the generator is represented exactly by the Hamiltonian matrix in Eq. (\ref{finite-rep}).
In the context of Koopman theory, the variational state $\ket{\psi_{\bm{\theta}}}$ can be regarded as a dictionary vector.
A dictionary vector is an infinite-dimensional vector whose entries are basis functions spanning a function space (e.g., $\theta_1,\theta_2,\theta_1\theta_2,\sin\theta_1,\cdots$); however, as discussed in Sec. \ref{sec4}, it is introduced in practice as a finite-dimensional approximation.
Except in special cases like the one considered here, a dictionary is generally not closed in a finite-dimensional space under nonlinear dynamics.
In data-driven Koopman analysis \cite{koopman1931hamiltonian,koopman1932dynamical,schmid2010dynamic, williams2015data,klus2020data,brunton2021modern}, the functional form of $\bm{f}(\bm{\theta})$ is typically unknown. Instead, a large number of data points of $(\bm{\theta}, \bm{f})$ are given, and the main task is to estimate the Koopman eigenvalues and eigenfunctions from these data.
By leveraging this, one can estimate the eigenvalues corresponding to the original Hamiltonian in a data-driven manner.
A schematic picture of the relationships among the quantum system, the nonlinear problem, and Koopman analysis is shown in Fig. \ref{fig1}(a).

\subsection{Sampling from low-dimensional variational manifold}
Next, consider a situation where the variational space covers only a small portion of the total Hilbert space. In quantum many-body physics, it is common to choose a variational space that can approximately describe the ground state.
In this situation, it is rare for the imaginary-time dynamics to close within the variational space [Fig. \ref{fig1}(b)].
Therefore, as sample points, we retain only values of $\bm{\theta}$ for which the residual is below a prescribed threshold.
In this paper, we use the following relative residual:
\begin{align}
    r(\bm{\theta})=\frac{\| \bm{f}(\bm{\theta})\cdot\nabla_{\bm{\theta}}\ket{\psi_{\bm{\theta}}}+H \ket{\psi_{\bm{\theta}}}\|}{\|H \ket{\psi_{\bm{\theta}}}\|}.\label{zansa}
\end{align}
For each $\bm{\theta}$, $\bm{f}$ is determined so as to minimize $r(\bm{\theta})$.
In the present case, $\bm{f}$ can be determined by performing a least-squares fit to the terms in the numerator.
The result is
\begin{align}
    &\bm{f}=-S^{-1}\bm{b},\label{least1}\\
    &S_{ij}=\bra{\partial_i \psi} \partial_j \psi\rangle,~b_i=\bra{\partial_i \psi}H\ket{\psi}.\label{least2}
\end{align}
Except for the normalization, this equation is the same as the one used in stochastic reconfiguration \cite{sorella2001generalized, sorella2005wave}.
At the sample points, the Koopman eigenequation (\ref{koopmaneigen}) is expected to hold approximately.
In this paper, we deliberately apply variational methods to systems that can be easily diagonalized, so that the evaluation of the residuals can be carried out straightforwardly. In contrast, in practical applications one must deal with very large-scale systems that genuinely require variational approaches; in such cases, whether a point can be used as a sample is determined based on residual estimates available for each variational method.
In Sec. \ref{section:mps}, we discuss the estimation of errors when using a uniform matrix product state as the variational wave function.

Now Eq. (\ref{nonlinear}) holds only for a limited set of sample points. Formally, one can consider a sufficiently large space containing the variational space, within which a continuous dynamics can be defined:
\begin{align}
    \dot{\bm{\Theta}}=\bm{F}(\bm{\Theta}).\label{truedynamics}
\end{align}
The corresponding Koopman eigenequation is given by
\begin{align}
    \bm{F}(\bm{\Theta})\cdot\nabla_{\bm{\Theta}}~z_i(\bm{\Theta})=-E_i~z_i(\bm{\Theta}).\label{largekoopman}
\end{align}
Conversely, the sample points can be thought of as embedded within a high-dimensional space, and they form an approximately closed dynamics within the embedded dimensions:
\begin{align}
    \bm{\Theta}_{\rm sample}=
    \begin{pmatrix}
    \bm{\theta}_{\rm sample}\\
    \bm{0}
    \end{pmatrix},~
    \bm{F}_{\rm sample}=
    \begin{pmatrix}
    \bm{f}_{\rm sample}\\
    \bm{0}
    \end{pmatrix}.\label{embedding}
\end{align}
Since Eq. (\ref{koopmaneigen}) holds approximately at sample points $\bm{\theta}_{\rm sample}$, Eq. (\ref{largekoopman}) also holds approximately at sample points $\bm{\Theta}_{\rm sample}$.
In other words, at the sample points, a trajectory of true dynamics (\ref{truedynamics}) is tangent to 
the variational space [Fig. \ref{fig1}(b)].

The above discussion means that there exists a true Koopman eigenequation specified by $\bm{F}$, and that the data points are sampled from a lower-dimensional variational space.
From a statistical perspective, there is a bias arising from restricting the data points to those within the variational space.
Therefore, the choice of variational space is expected to affect the accuracy of the eigenvalue predictions. For example, if one is interested in the ground-state energy, it is desirable to design a variational space such that the sample points are more densely distributed in the invariant subspace corresponding to low-energy dynamics.

In this paper, we use the following type of variational wave functions:
\begin{align}
    \ket{\psi_{\bm{\theta}}}=\theta_{0}\ket{\tilde{\psi}_{\bm{\theta}^{(d)}}},\label{varinpaper}
\end{align}
where $\bm{\theta}^{(d)}=(\theta_1,\theta_2,\cdots,\theta_d)$,  $\bm{\theta}=(\theta_0,\bm{\theta}^{(d)})$, and $\ket{\tilde{\psi}_{\bm{\theta}^{(d)}}}$ is a nonzero vector.
By introducing the overall factor $\theta_{0}$, the variational wave function is able to capture changes in the norm, thereby expanding the range of data points for which the residual falls below the threshold.
By construction, $\theta_{0}=0$ corresponds to the stable fixed-point hyperplane of the nonlinear dynamics (\ref{nonlinear}).
If the Hamiltonian is represented by a complex-valued matrix, it is preferable to introduce one additional variational parameter to account for the phase degree of freedom. In this paper, we assume that all Hamiltonians under consideration are real.
In the following sections, we first consider analytically tractable examples with sufficiently large variational spaces in order to gain insight into the application of Koopman theory.
We then turn to the main target of this study: cases where the variational space is restricted and must be addressed using data-driven approaches.
\section{Analytic example: Two-level system\label{sec3}}
As an example that provides an analytical understanding of the relationship between a quantum system, a nonlinear system, and the Koopman generator, we consider a two-level system:
\begin{align}
    H=
    \begin{pmatrix}
    \lambda_1&0\\
    0&\lambda_2
    \end{pmatrix},
\end{align}
where $0<\lambda_1<\lambda_2$.
As a variational state, we consider
\begin{align}
    \ket{\psi_{\bm{\theta}}}=\theta_{0}\frac{1}{\sqrt{1+\theta_1^2}}
    \begin{pmatrix}
    1\\
    \theta_1
    \end{pmatrix},\label{varfunc}
\end{align}
where $\bm{\theta}\in\mathbb{R}^2$.
This variational space is evidently large enough to represent the dynamics of this real-valued Hamiltonian.
Therefore, in this example, the dynamics (\ref{variationalsch}) or (\ref{finite-rep}) is exactly closed for any $\bm{\theta}$ without the need to compute the residual.
In this simplest case, the Koopman eigenvalues and eigenfunctions corresponding to the original Hamiltonian can be obtained by substituting Eq. (\ref{varfunc}) into Eq. (\ref{variationalsch}), as follows:
\begin{align}
\dot{\bm{\theta}}\cdot\nabla_{\bm{\theta}}z_1(\bm{\theta})=-\lambda_1 z_1(\bm{\theta}),~\dot{\bm{\theta}}\cdot\nabla_{\bm{\theta}}z_2(\bm{\theta})=-\lambda_2 z_2(\bm{\theta}),
\end{align}
where
\begin{align}
    z_1(\bm{\theta})=\frac{\theta_0}{\sqrt{1+\theta_1^2}},~z_2(\bm{\theta})=\frac{\theta_0\theta_1}{\sqrt{1+\theta_1^2}}.
\end{align}
In addition, the analytical expression of the nonlinear function $\bm{f}$ is calculated by taking the time derivatives of $z_1$ and $z_2$, and the following holds:
\begin{align}
    \dot{\bm{\theta}}=\bm{f}(\bm{\theta})=-
    \begin{pmatrix}
    \lambda_1 ~\theta_0\\
    (\lambda_2-\lambda_1)~\theta_1
    \end{pmatrix}
    -
    \begin{pmatrix}
    (\lambda_2-\lambda_1)\frac{\theta_1^2}{1+\theta_1^2}\\
    0
    \end{pmatrix}.
\end{align}
In this manner, the nonlinear system corresponding to a given quantum Hamiltonian, as well as its Koopman generator, are computed analytically.

In general, it is preferable for the variational state to have an analytically simple form.
For example, let us consider the same problem without the normalization factor $\sqrt{1+\theta_1^2}$.
Repeating the above argument, one obtains the corresponding Koopman eigenfunctions, $\tilde{z}_1(\bm{\theta})=\theta_0$ and $\tilde{z}_2(\bm{\theta})=\theta_0\theta_1$, and the $\bm{\theta}$ dynamics is given by a linear dynamics:
\begin{align}
    \dot{\bm{\theta}}=\tilde{\bm{f}}(\bm{\theta})=-
    \begin{pmatrix}
    \lambda_1 ~\theta_0\\
    (\lambda_2-\lambda_1)~\theta_1
    \end{pmatrix}.
\end{align}
Note that, regardless of which variational function is used, $\theta_1$ follows a linear dynamics with eigenvalue $-(\lambda_2-\lambda_1)$. 
By setting $z(\bm{\theta})=\theta_1$, we see that this eigenvalue is also a Koopman eigenvalue.  
As in this case, the set of Koopman eigenvalues includes eigenvalues that are not related to the eigenvalues of the original Hamiltonian.
Note also that, because the product of eigenfunctions of a first-order differential operator is also an eigenfunction, the sum of Koopman eigenvalues is again a Koopman eigenvalue. In the above example, the following relation holds:
\begin{align}
    -\lambda_2=-\lambda_1+\left[-(\lambda_2-\lambda_1)\right].
\end{align}
As can be seen from the above, it is important to identify which Koopman eigenvalues correspond to those of the original quantum system.
From the assumption in Eq.~(\ref{varinpaper}), the associated eigenfunction is at least proportional to $\theta_0$.

\section{Data-driven Koopman analysis for 4-site transverse-field Ising model\label{sec4}}
In the two-level example above, the variational space was sufficiently large to capture the full dynamics of the Hamiltonian; however, in practice, we are interested in cases where the variational space is smaller than the Hilbert space.
In this section, we take the transverse-field Ising model as an example and perform spectral analysis using a finite-dimensional approximation of the Koopman generator known as extended dynamic mode decomposition (EDMD) \cite{williams2015data, klus2020data}.

\subsection{Model and variational state}
We consider the one-dimensional transverse-field Ising model with periodic boundary condition \cite{kitaev2010topological,fradkin2013field}:
\begin{align}
    \hat{H}=-J\sum_{i}\sigma^z_{i}  \sigma^z_{i+1}-h\sum_{i}\sigma^x_i+C,
\end{align}
where $\sigma^{x,z}_i$ are the Pauli matrices at site $i$, and a constant term $C$ is added to ensure positive definiteness. We assume $J,h>0$ for simplicity.
For $h>J$, the spins are strongly influenced by the transverse magnetic field, and the ground state tends to align along the $x$ direction. The system has a unique, gapped ground state. This phase is called the field-polarized phase, also known as the disordered phase.
For $h<J$, due to the spin-spin interaction, the ground state becomes twofold degenerate, corresponding to the ferromagnetic phase, also referred to as the ordered phase.
While both of these phases are gapped, at $h=J$, the system undergoes a quantum phase transition, at which the Hamiltonian exhibits gapless excitations.

In the following, we consider the disordered phase.
As is well known, this model is exactly solvable \cite{kitaev2010topological,fradkin2013field}. However, we intentionally consider a variational subspace that excludes the ground state.
For convenience, we introduce the number basis $\ket{0}_i:=\ket{\sigma^x_i=+1}$ and $\ket{1}_i:=\ket{\sigma^x_i=-1}$ with the hardcore boson operators of ``magnons":
\begin{align}
    a_i^{\dagger}\ket{0}_i=\ket{1}_i,~a_i\ket{1}_i=\ket{0}_i.
\end{align}
As a variational state, we use the following translation-invariant state:
\begin{align}
    \ket{\psi_{\bm{\theta}}}=\exp\left(\sum_{i>j}\theta(|i-j|)~ a^{\dagger}_ia^{\dagger}_j\right)\left[\bigotimes_i\ket{0}_i\right],
\end{align}
where $\theta(|i-j|)$ is the distance-dependent variational parameter.
This state contains only configurations with an even number of bosons, and the Hamiltonian acts within each boson-number-parity sector.
Here, we consider a four-site system restricted to the even boson-number sector.
Because of the periodic boundary condition, the 4-site system has only two independent parameters:
\begin{align}
    &\theta_1:=\theta(|i-j|=1)=\theta(|i-j|=3),\notag\\
    &
    \theta_2:=\theta(|i-j|=2).
\end{align}
Since the dimension of the Hamiltonian matrix within the even-boson-number sector is $D=2^4/2=8$, the dynamics does not close within the two-parameter variational space. 
Hence, this case provides a minimal example suitable for data-driven Koopman analysis.
We set $J=1$, $h=2$, and $C=9$.
The six lowest energy eigenvalues in both sectors are given by 0.45688, 2.52786, 5, 5
,6.35139, and 6.52786.

\subsection{Preparation of Dataset}

\begin{figure}[]
\begin{center}
 \includegraphics[width=7cm,angle=0,clip]{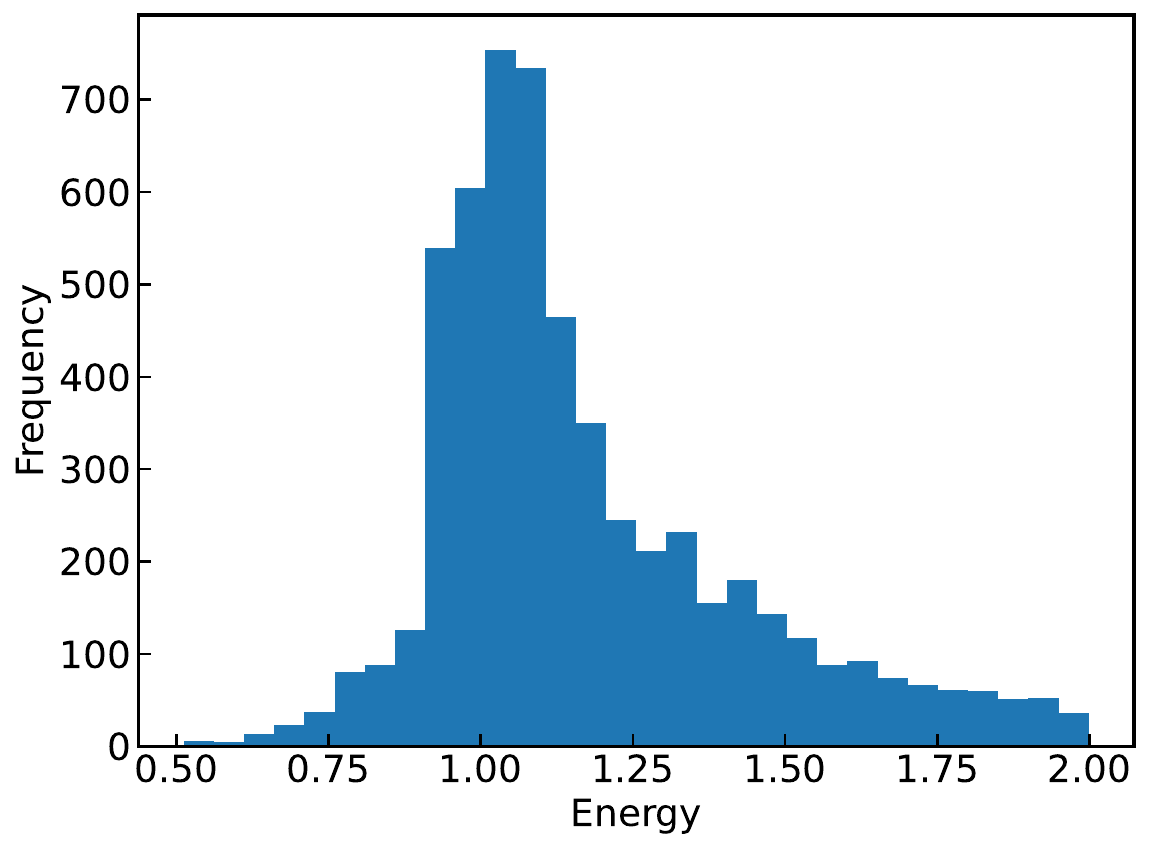}
 \caption{Histogram of sample energies.}
 \label{fig2}
\end{center}
\end{figure}

\begin{figure}[]
\begin{center}
 \includegraphics[width=8cm,angle=0,clip]{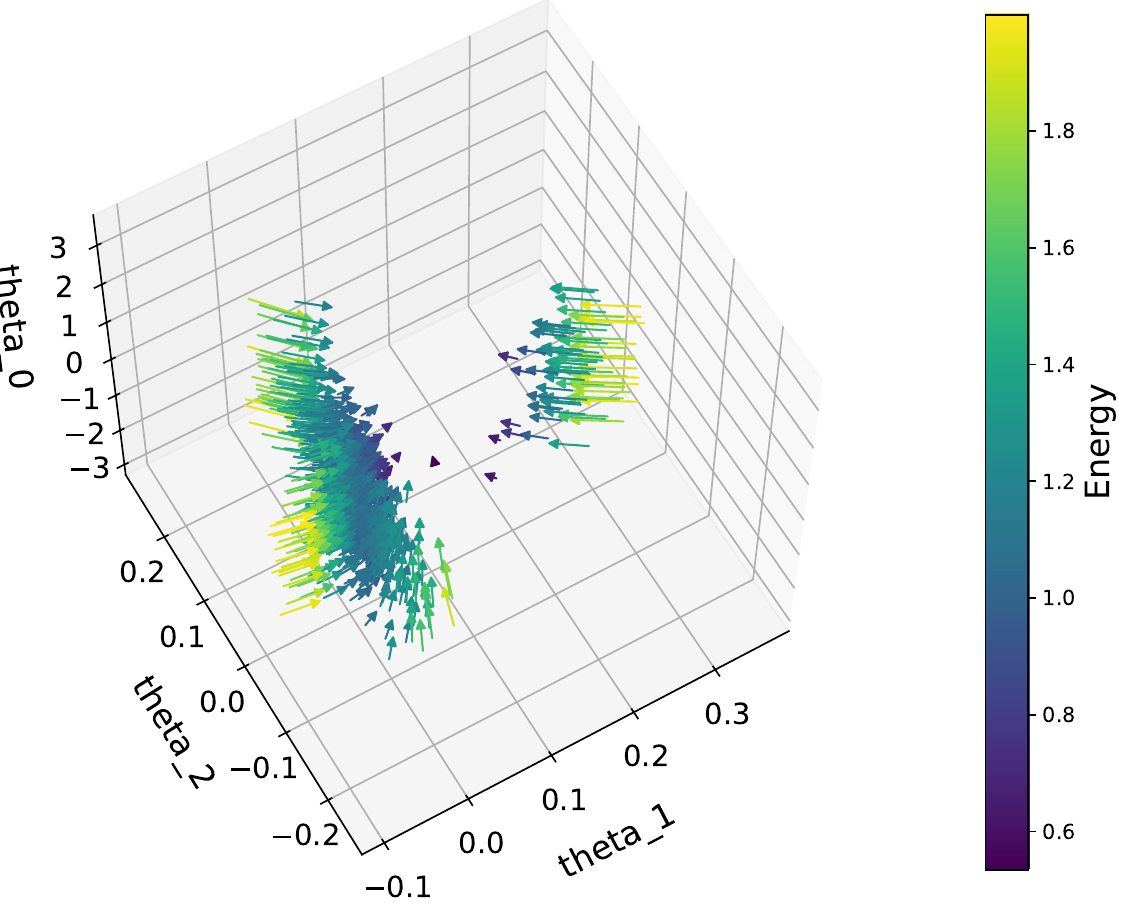}
 \caption{Vector field plot of $\bm{f}(\bm{\theta})$. The color of the arrows represents the expectation value of the sample energy.}
 \label{fig3}
\end{center}
\end{figure}

A single sample point is generated by the following procedure with a fixed random seed.
First, the initial value of $\bm{\theta}$ is generated randomly such that each component  $\theta_i$ follows a standard normal distribution $\mathcal{N}(0,1)$.
Starting from the initial value generated in this manner, $\bm{\theta}$ is optimized until the relative residual (\ref{zansa}) becomes smaller than the threshold $10^{-3}$ or until a maximum of 100 iterations is reached.
In the optimization, $\bm{f}$ is first computed by performing a least-squares fit (\ref{least1}) and (\ref{least2}) with $\bm{\theta}$ fixed, and then $\bm{\theta}$ is updated using a gradient-based method.
The above procedure is performed sequentially for seeds starting from $0$, and only the runs that reach the threshold are retained as samples. This yields $N_{\rm sample}$ pairs of $(\bm{\theta},\bm{f})$, each satisfying Eq. (\ref{finite-rep}) within the specified tolerance.
To facilitate the prediction of the ground-state energy, samples with excessively high energies are excluded. Here, we consider 5,689 samples with energy values less than or equal to 2 (Fig. \ref{fig2}).
Figure \ref{fig3} shows a vector plot of 1,000 randomly selected sample points.
As shown in Fig. \ref{fig2}, the energy distribution of the samples is not uniform, and its effect on the analysis is nontrivial. In this study, we use the non-uniform data as is; however, for more advanced analyses, it may be necessary to take the influence of the distribution into account.

\subsection{Extended dynamic mode decomposition (EDMD)}
\begin{figure}[]
\begin{center}
 \includegraphics[width=8cm,angle=0,clip]{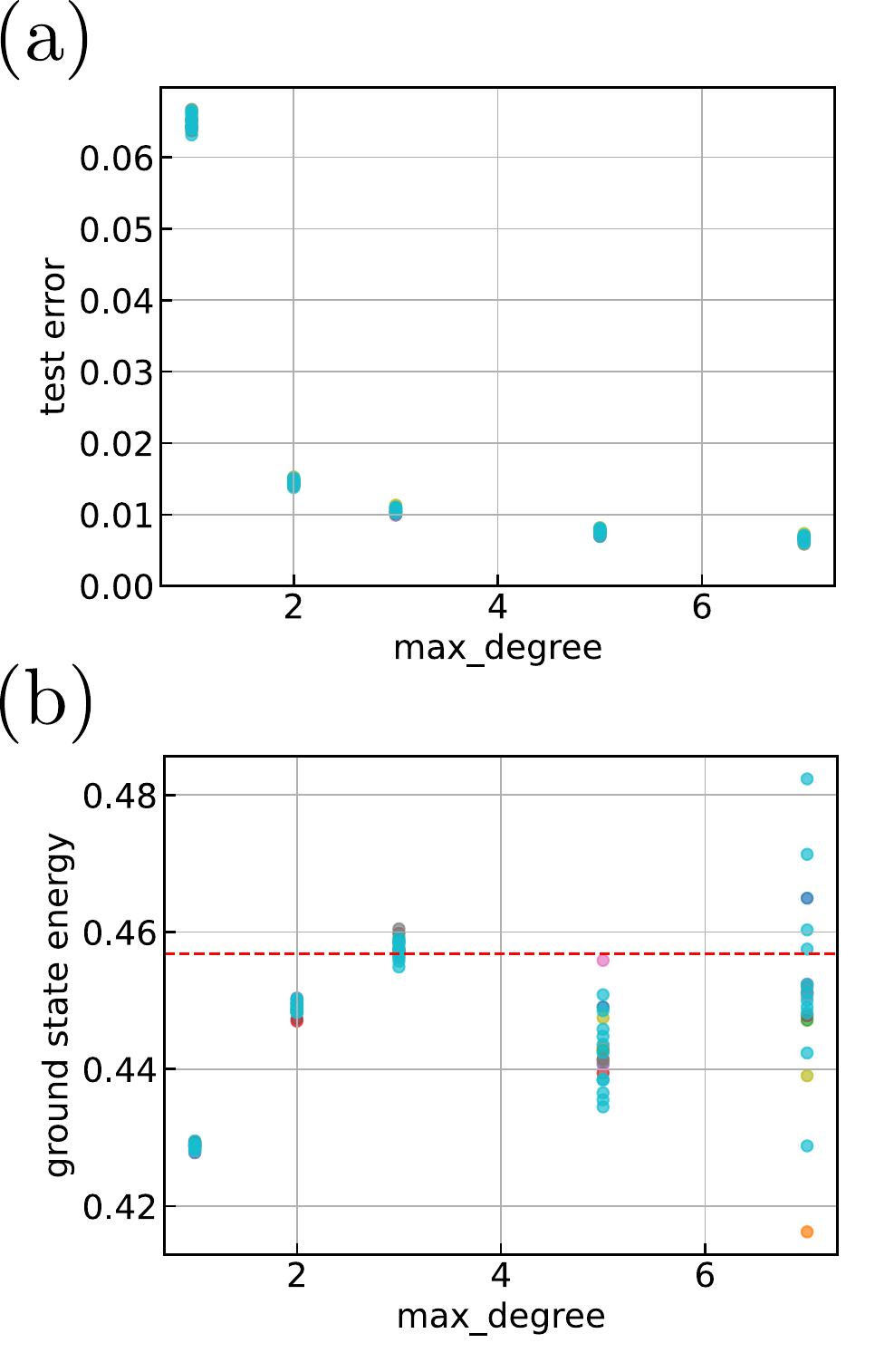}
 \caption{Distributions of (a) the test error and (b) the estimated ground-state energy as functions of the maximum degree of the monomial dictionary. The red dashed line represents the true ground-state energy, 0.45688. }
 \label{fig4}
\end{center}
\end{figure}
In the following, we perform EDMD \cite{williams2015data, klus2020data} for the Koopman generator, also referred to as gEDMD \cite{klus2020data}.
In EDMD, the Koopman generator, which is defined on an infinite-dimensional Hilbert space of functions, is approximated by projecting it onto a finite-dimensional function subspace spanned by a prescribed set of basis functions.
Our task is to find a finite-dimensional matrix approximation of the Koopman generator:
\begin{align}
    \frac{d}{d\tau}\ket{\phi_{\bm{\theta}}}\simeq L_{N}\ket{\phi_{\bm{\theta}}}~\mathrm{with~}\frac{d}{d\tau}=\bm{f}(\bm{\theta})\cdot\nabla_{\bm{\theta}}~,
\end{align}
where $L_N$ is a $N\times N$ matrix.
The dictionary vector $\ket{\phi_{\bm{\theta}}}$ contains the basis functions as its elements.
Considering the same expression for all data points and collecting them together, we can write this in matrix form as follows:
\begin{align}
    \dot{\Phi}\simeq L_N~\Phi,\label{finite-dim-approx}
\end{align}
where $\Phi$ and $\dot{\Phi}$ are $N\times N_{\rm sample}$ non-square matrices defined as
\begin{align}
    &\Phi:=\left(\ket{\phi_{\bm{\theta}_1}},\ket{\phi_{\bm{\theta}_2}},\cdots,\ket{\phi_{\bm{\theta}_{N_{\rm sample}}}}\right),\\
    &\dot{\Phi}:=\left(\frac{d}{d\tau}\ket{\phi_{\bm{\theta}_1}},\frac{d}{d\tau}\ket{\phi_{\bm{\theta}_2}},\cdots,\frac{d}{d\tau}\ket{\phi_{\bm{\theta}_{N_{\rm sample}}}}\right).
\end{align}
Note that, in order to better align with the Schr\"{o}dinger equation, we adopt transposed definitions of the usual $\Phi$, $\dot{\Phi}$, and $L_N$.
By performing linear regression on Eq. (\ref{finite-dim-approx}), $L_N$ is estimated as
\begin{align}
    L_N=\dot{\Phi}\Phi^+,
\end{align}
where $+$ denotes the Moore–Penrose pseudoinverse.
Numerically, Eq. (\ref{finite-dim-approx}) is solved using a standard linear solver.
In order to obtain numerically stable results, it is necessary that $N_{\rm sample}\gg N$.
As can be seen from this construction, $L_N$ is not necessarily Hermitian.
Note that, when the non-normality, measured by the commutator $[L_N,L_N^{\dagger}]$, is large, the eigenvalues can become unstable. This non-normality originates both from artificial factors—such as the non-orthogonality of the dictionary and the non-uniformity of the samples—and from intrinsic nonlinear effects, including chaos.

This formulation shows that the choice of the dictionary vector can significantly affect the results.
Clearly, by taking $\ket{\phi_{\bm{\theta}}}=\ket{\psi_{\bm{\theta}}}$ and $L_N=-H$, the imaginary-time Schr\"{o}dinger equation (\ref{variationalsch}) is recovered.
In this sense, the variational state itself can be regarded as a dictionary vector, as mentioned before.
In practice, $\ket{\psi_{\bm{\theta}}}$ is a very high-dimensional vector in the original quantum system and cannot be used directly as a dictionary vector.
Instead, by ensuring that the dictionary vector preserves the property of being proportional to $\theta_0$, we can maintain a good correspondence with the original quantum system.
We use a monomial basis whose elements are proportional to $\theta_0$.
In the case where $\theta_1$ and $\theta_2$ are linear, the dictionary vector is given by a 3-dimensional vector:
\begin{align}
    \ket{\phi_{\bm{\theta}}}=\theta_{0}
    \begin{pmatrix}
    1\\
    \theta_1\\
    \theta_2
    \end{pmatrix}.
\end{align}
Below, we perform EDMD using dictionaries of various degrees in $\theta_1$ and $\theta_2$, and estimate the smallest eigenvalue of $-L_N$, which corresponds to the ground-state energy.
Half of the samples are used for training, and the remaining half are used to estimate the test error. The relative error is defined as follows:
\begin{align}
    r=\frac{\| \dot{\Phi}-L_N\Phi\| _F}{\| \dot{\Phi}\|_F},
\end{align}
where $\|\cdot\|_F$ denotes the Frobenius norm.
When using a monomial dictionary, the non-normal nature of $-L_N$ becomes more pronounced as the degree increases, making the eigenvalues increasingly unstable. Taking this into account, we consider 20 random splits of the samples and compute the smallest eigenvalue for each case.
To ensure consistent scaling across different monomial degrees, each row of $\Phi$ is normalized for the training data, and the same scaling is applied to the test data.
Figure \ref{fig4}(a,b) shows the test error and the smallest eigenvalue for the cases where the degrees of $\theta_1$ and $\theta_2$ are 1, 2, 3, 5, and 7. The cases with degrees 4 and 6 are excluded from the plots because the smallest eigenvalue becomes highly unstable. As seen in Fig. \ref{fig4}(a), the test error improves as the degree increases. This indicates that the number of samples is sufficient and that overfitting is well suppressed. On the other hand, as shown in Fig. \ref{fig4}(b), the instability of the eigenvalues becomes significantly more pronounced when the degree exceeds 3. The case of degree 3, which achieves a good balance between test error and eigenvalue stability, reproduces the true eigenvalue 0.45688 quite well. Meanwhile, even the linear case provides a reasonably good prediction, suggesting that the nonlinearity of the dynamics is relatively weak. This is likely because we only considered samples with sufficiently low energy.

\section{Application for time-dependent variational principle for normalized matrix product states\label{section:mps}}
So far, we have considered cases where the Hamiltonian matrix is small enough to be diagonalized. However, the situations we truly want to address in applications are precisely those where the matrix is too large for direct diagonalization. To obtain samples in such cases, it is necessary to evaluate the errors of the imaginary-time dynamics and the quantity $\bm{f}$ without explicitly using the full large matrix. As an application, we discuss how to apply Koopman theory within the framework of the time-dependent variational principle (TDVP \cite{dirac1930note,langhoff1972aspects}) for normalized uniform matrix product states \cite{haegeman2011time,haegeman2013post,haegeman2016unifying,vanderstraeten2019tangent}.

For a normalized state $\ket{\phi}=\ket{\psi}/\sqrt{\bra{\psi}\psi\rangle}$, the imaginary-time dynamics is given, as recalled here, by
\begin{align}
    \frac{d}{d\tau}\ket{\phi}=-\left(H-\langle H\rangle_{\phi}\right)\ket{\phi},\label{app2}
\end{align}
where $\langle H\rangle_{\phi}:=\bra{\phi}H\ket{\phi}$.
Then, Eq. (\ref{imgeqn}) is rewritten as
\begin{align}
    \frac{d}{d\tau}\ket{\psi}=&-\langle H\rangle_{\phi}\bra{\psi}\psi\rangle^{1/2}\ket{\phi}\notag\\&-\bra{\psi}\psi\rangle^{1/2}\left(H-\langle H\rangle_{\phi}\right)\ket{\phi}.\label{non-normalized-true}
\end{align}

Next, we consider the TDVP for the uniform matrix product state $\ket{\phi (A)}$ in the infinite-volume limit \cite{haegeman2011time}:
\begin{align}
    \ket{\phi (A)}=\sum_{\{s\}}\bm{v}^{\dagger}_L\left[\prod_{m\in\mathbb{Z}}A^{s_m}\right]\bm{v}_R\ket{\{s\}},
\end{align}
where $s$ denotes local degrees of freedom such as spin, $\ket{\{s\}}$ is a physical basis, $A^s$ is a $\chi\times\chi$ matrix, and $A$ denotes the tensor whose elements are given by $[A^s]_{ab}$. 
The $\chi$-dimensional vectors $\bm{v}_{L,R}$ are usually irrelevant in the infinite-volume limit. 
The $A$ tensor is chosen such that the state is normalized. Here, the variational parameters are the elements of the $A$ tensor.
The TDVP is a framework for obtaining approximate quantum dynamics by replacing the action of the Hamiltonian at each time step with the best possible state within the variational manifold.
In particular, for matrix product states, efficient methods based on the tangent space are known \cite{haegeman2013post,haegeman2016unifying,vanderstraeten2019tangent}. In these approaches, the action of the Hamiltonian on $\ket{\phi (A)}$ is approximated by projecting it onto the tangent space of $\ket{\phi (A)}$. As a result, even in the case of imaginary-time dynamics, the norm is conserved.
The TDVP imaginary-time dynamics is given by \cite{haegeman2013post,haegeman2016unifying,vanderstraeten2019tangent}
\begin{align}
    \frac{d}{d\tau}\ket{\phi(A)}&=-P_{\rm TDVP}~H\ket{\phi(A)}\notag\\
    &=-P_{\rm TDVP}~\left(H-\langle H\rangle_{\phi}\right)\ket{\phi(A)},
\end{align}
where $P_{\rm TDVP}$ is the projection operator onto the tangent space.
In the second line, we have used $P_{\rm TDVP}~\ket{\phi(A)}=0$.
The error from the true dynamics (\ref{app2}) is given by
\begin{align}
    \epsilon=\|\left(1-P_{\rm TDVP}\right) \left(H-\langle H\rangle_{\phi}\right)\ket{\phi(A)}  \|.
\end{align}
Notably, this quantity (per site) can be computed efficiently \cite{haegeman2013post}.
Similarly, since $\|P_{\rm TDVP}~\left(H-\langle H\rangle_{\phi}\right)\ket{\phi(A)}\|$ can also be computed efficiently, we can define the following relative residual, which allows us to estimate how closely a given $\ket{\phi(A)}$ follows the true dynamics:
\begin{align}
    r(A)&=\frac{\|\left(1-P_{\rm TDVP}\right) \left(H-\langle H\rangle_{\phi}\right)\ket{\phi(A)}  \|}{\|P_{\rm TDVP}~\left(H-\langle H\rangle_{\phi}\right)\ket{\phi(A)}\|}\notag\\
    &=\frac{\|\left(1-P_{\rm TDVP}\right) \left(H-\langle H\rangle_{\phi}\right)\ket{\phi(A)}  \|}{\|P_{\rm TDVP}~H\ket{\phi(A)}\|}.\label{mpsresidual}
\end{align}
We use only the $A$ tensors with a small relative residual as samples.

Finally, we apply the TDVP dynamics to a non-normalized variational state $\ket{\psi_{\bm{\theta}}}=\theta_0\ket{\phi(A)}$.
The dynamics of $\ket{\psi_{\bm{\theta}}}$ is given by
\begin{align}
\frac{d}{d\tau}\ket{\psi_{\bm{\theta}}}=\dot{\theta}_0\ket{\phi(A)}-\theta_0P_{\rm TDVP}~\left(H-\langle H\rangle_{\phi}\right)\ket{\phi(A)}.
\end{align}
By setting $\theta_0=\bra{\psi}\psi\rangle^{1/2}$ and $\dot{\theta}_0/\theta_0=-\langle H\rangle_{\phi}$, we find that this equation provides an approximation of the true dynamics of a non-normalized state, (\ref{non-normalized-true}). 
The error arises only from the second term and is given by $\theta_0\epsilon$.
Thus, the relative residual for $\ket{\psi_{\bm{\theta}}}$ is also given by Eq. (\ref{mpsresidual}).
By setting $z(\bm{\theta})=\theta_0\tilde{z}(A)$, we obtain the ``modified" eigenequation for the Koopman generator:
\begin{align}
    &\frac{d}{d\tau}z(\bm{\theta})=-\lambda z(\bm{\theta})\notag\\
    \Leftrightarrow&\frac{d}{d\tau}\tilde{z}(A)+\frac{\dot{\theta_0}}{\theta_0}\tilde{z}(A)=-\lambda\tilde{z}(A)\notag\\
    \Leftrightarrow&\bm{f}(A)\cdot\nabla_{A}\tilde{z}(A)=-(\lambda-\langle H\rangle_A)\tilde{z}(A).
\end{align}
Unlike the formalism with a non-conserved norm, in this formalism, a shift of the energy zero does not affect the dynamics.
In addition, the overall factor $\theta_0$ is removed from the theory.
For a uniform matrix product state, $\langle H\rangle_A:=\bra{\phi(A)}H\ket{\phi(A)}$ and $\bm{f}$ for a given matrix $A$ are efficiently calculated.
By definition, $\bm{f}$, $\lambda$, and $\langle H\rangle_A$ are proportional to the volume. In practice, we divide both sides by the volume to obtain an equation for the corresponding densities.
The corresponding EDMD equations are modified as follows:
\begin{align}
    &\dot{\Phi}'\simeq L_N~\Phi,\\
    &\dot{\Phi}':= \dot{\Phi}-\Phi_{E},\\
    &\Phi_E:=\left(\cdots,\langle H\rangle_A~\ket{\phi(A)},\cdots\right).
\end{align}
Here, $\ket{\phi}$ represents a general dictionary vector rather than a matrix product state.

Note that the computation of $\bm{f}$ can involve the inversion of an ill-conditioned matrix. Such a situation occurs when there are small Schmidt coefficients.
This is a well-known issue in the TDVP formulation for uniform matrix product states. It can be overcome by transforming the uniform matrix product state into the mixed canonical form \cite{haegeman2016unifying}. Because this method involves discrete updates with small time steps, it may be more convenient to introduce a discrete-time Koopman analysis.
For a discrete nonlinear process $\bm{\theta}_{n+1}=T[\bm{\theta}_n]$, the Koopman operator $K$, in a matrix representation, is defined as
\begin{align}
    \ket{\Phi(T[\bm{\theta}_n])}=K\ket{\Phi(\bm{\theta}_n)},
\end{align}
where $\ket{\Phi}$ is an infinite-dimensional vector.
Then, a Koopman eigenstate obeys
\begin{align}
    z_i(T[\bm{\theta}_n])=\mu_iz_i(\bm{\theta}_n),
\end{align}
where $\mu_i$ is a discrete Koopman eigenvalue.
In a data-driven manner, the set $\{\bm{\theta}_n,\bm{\theta}_{n+1}\}$ corresponds to one sample.
Actually, methods for the discrete case have been more extensively studied. In particular, there exist prior works on EDMD approaches that learn the dictionary using neural networks, as well as autoencoder-based methods that learn both the dictionary and the Koopman operator \cite{brunton2021modern,yeung2019learning,lusch2018deep,li2017extended,takeishi2017learning,otto2019linearly}.
In our problem with norm conservation, these equations correspond to
\begin{align}
    \ket{\phi(T[A])}&=e^{\delta\tau ~(L_N+\langle H\rangle_{A})}\ket{\phi(A)},\\
    \tilde{z}(T[A])&=e^{\delta \tau~(\lambda+\langle H\rangle_{A})}\tilde{z}(A).
\end{align}
Since a matrix product state is invariant under similarity transformations of the matrices $A^s$, one must either fix an appropriate gauge or analyze 
$\tilde{z}$ in a gauge-invariant way.
\section{Discussion\label{discussion}}
This section discusses several remaining issues.
The most straightforward application of the present method, apart from the matrix product states discussed in the main part, would be to the TDVP in variational Monte Carlo \cite{foulkes2001quantum}. In variational Monte Carlo, one minimizes the energy expectation value within a variational manifold, and the stochastic reconfiguration method \cite{sorella2001generalized, sorella2005wave}—proposed as an efficient approach for this purpose—can be interpreted as the TDVP applied to the imaginary-time Schrödinger equation. In this sense, the quantity $\dot{\bm{\theta}}$ at a given $\bm{\theta}$ is a standard quantity computed in variational Monte Carlo. On the other hand, compared to this quantity, the explicit evaluation of the dynamical error itself is less frequently performed. Such errors have been discussed, for example, in Ref. \cite{takai2016finite} for the imaginary-time case and in Refs. \cite{carleo2017solving, schmitt2020quantum} for the real-time case. Using these criteria, one can assess whether a given $\bm{\theta}$ is suitable as a sample for Koopman analysis.

As another issue, the influence of sampling on estimating the eigenvalue structure, including the ground state, is of particular interest. As already discussed, the space from which samples are drawn is a highly restricted subset of the variational space, which itself is only a small portion of the full Hilbert space—namely, the subspace where the dynamical error is small. In this sense, our problem can be regarded as an eigenvalue estimation problem for a Koopman generator with a non-uniform sampling measure.
The present method may become advantageous over existing variational approaches in situations where the ground state is not contained within the variational space. In such cases, it is nontrivial to what extent the eigenvalue structure can be accurately predicted. Developing more sophisticated analysis methods, including those based on neural networks, remains an important direction for future work.

\acknowledgements
I thank Takahiro Misawa, Kota Ido, and Yusuke Nomura for kindly providing helpful information and references on the variational Monte Carlo method.
This work was supported by JSPS KAKENHI Grant No.~JP23K03243.

\bibliography{Koopman}
\end{document}